\documentclass[prd,aps,floatfix,nofootinbib,preprint ,tightenlines,superscriptaddress]{revtex4}
\usepackage{dcolumn}
\usepackage{amsmath}
\usepackage{latexsym}
\usepackage{graphicx}
\usepackage{bm}

\usepackage{tikz-feynman}

\def\svev#1{\left\langle #1\right\rangle}       

\def\Tr{{\rm Tr}\,}

\newcommand{\bee}{\begin{equation}}
\newcommand{\ee}{\end{equation}}
\newcommand{\beea}{\begin{eqnarray}}
\newcommand{\eea}{\end{eqnarray}}

\tikzfeynmanset{compat=1.1.0}

\begin{document}
\title{Comparison of a pseudoscalar  meson form factor in QCD with 3, 4, and 5 colors}
\author{Thomas DeGrand}
\affiliation{Department of Physics,
University of Colorado, Boulder, CO 80309, USA}
\email{thomas.degrand@colorado.edu}

\date{\today}

\begin{abstract}
I show comparisons of the pseudoscalar meson vector form factor 
from simulations of QCD with $N_c = 3$, 4 and 5 colors
and $N_f = 2$ flavors of degenerate mass fermions at a common (matched) fermion mass, lattice spacing,
and simulation volume.
The dependence of the form factor on the momentum transfer is nearly independent of
 the number of colors, and is consistent with the expectations of vector meson dominance.
\end{abstract}
\maketitle

\section{Introduction}

It is generally expected
 \cite{tHooft:1973alw,tHooft:1974pnl,Witten:1979kh}
that as the number of colors $N_c$ becomes very large,  the mesonic sector of QCD is described
by planar amplitudes, interacting open strings with quarks and antiquarks
on their perimeters \cite{Harari:1969oxx,Rosner:1969bhr}. This says further that properties of mesons
are independent of $N_c$ up to overall  (process dependent) $N_c$ scaling factors, at least
to leading order in $N_c$. The parameter which connects results with different numbers of colors is
the 't Hooft coupling $\lambda=g^2 N_c$.

Lattice simulations have tested much of this paradigm.
(For reviews, see Refs.~\cite{Lucini:2012gg,GarciaPerez:2020gnf,Hernandez:2020tbc}.)
The bare 't Hooft coupling sets the lattice spacing.
The overall scaling $N_c$ has been seen many times, in (for example) decay constants
of pseudoscalar and vector mesons  or the kaon $B$ parameter.  The underlying independence of observables after
the $N_c$ dependence is removed has many indirect tests: one is how a quantity scales with fermion mass,
such as any hadron mass versus quark mass $m_q$.
 In quark models, decay constants are proportional to a color factor (typically a power of $N_c$) times
the squared quark wave function at zero separation $|\psi(0)|^2$  (see for example Ref.~\cite{VanRoyen:1967nq}), and seeing the
appropriate overall $N_c$ scaling is a statement that $|\psi(0)|^2$ is independent of $N_c$.

It would be nice to have a more direct comparison of the underlying independence of mesonic matrix elements across $N_c$.
This little note makes that comparison in  the vector form factor $F(q^2)$ of a pseudoscalar meson as a 
function of (spacelike) momentum transfer $q$. The form factor is defined as
\bee
\svev{\pi(p') | J_\mu(q) | \pi(p)} = (p'_\mu + p_\mu)F(q^2)
\label{eq:form}
\ee
where $J_\mu$ is the vector (electromagnetic) current. Current conservation requires that $F(0)=1$ so the shape is
the interesting quantity.
 In a quark model, again, the form factor is the Fourier transform
of the charge density of the meson and the goal is to see whether the shape of $F(q^2)$ varies with $N_c$.
I did an absolutely minimalist calculation: I looked at $N_c=3$, 4, and 5 systems at a single
lattice spacing, a single volume (chosen to avoid finite volume
artifacts) and a single quark mass, matched across $N_c$. In physical units, the lattice spacing is
roughly  0.1 fm, the lattices have a spatial size of 1.6 fm, and the pseudoscalar and vector meson 
masses are about 650 and 1050 MeV, respectively. (Lattice numbers will be given below, in Table \ref{tab:params}.)
The result is easy to state: over the range of momenta where I had a signal, the shape of the form factor
is independent of $N_c$.

There are two technical issues in the calculation. The first is that it
is necessary to have
good interpolating fields to produce
particles at nonzero three momentum. The second problem is that lattice calculations 
 produce the form factor along with other nuisance quantities, which must be separated in the
analysis.
People have been computing the form factor of the
pseudoscalar meson since at least 1987 (see Ref.~\cite{Martinelli:1987bh}) and by
now there are probably hundreds of papers about hadronic form factors in the literature with
  many methods  to deal with these problems.  One only has to search the literature and adapt something
  already there. I chose to  use the interpolating field
techniques of Refs.~\cite{Bali:2016lva} and \cite{Izubuchi:2019lyk}
and the fitting procedure of Ref.~\cite{FermilabLattice:2022gku}.

And a remark before we continue:
the goal of a little study like this is to ask whether versions of QCD with different numbers of colors
(but everything else fixed) produce different  values for an observable. The overall $N_c$ counting and dependence
of the quark mass is what is important.
High precision, chiral extrapolations, or a continuum limit can come later (if ever). 

So let us proceed. Sec.~\ref{sec:method} goes through all the technical aspects of the lattice calculation.
Results are found in Sec.~\ref{sec:results}. Sec.~\ref{sec:summary} summarizes and concludes.

\section{Methodology\label{sec:method}}
\subsection{Generating the data sets}
All my publications about large $N_c$ QCD used the same lattice
action and methodology.
A long description of techniques can be found  in Ref.~\cite{DeGrand:2023hzz}. Here is a short summary:

The gauge action  is the
 Wilson plaquette action.
Two flavors of degenerate mass  Wilson-clover fermions are simulated.
Configurations are generated using
 the Hybrid Monte Carlo (HMC)  algorithm \cite{Duane:1986iw,Duane:1985hz,Gottlieb:1987mq}
with a multi-level Omelyan integrator \cite{Takaishi:2005tz} and
multiple integration time steps \cite{Urbach:2005ji}
with one level of mass preconditioning for the fermions \cite{Hasenbusch:2001ne}.

The fermion action uses normalized hypercubic
 smeared links~\cite{Hasenfratz:2001hp,Hasenfratz:2007rf,DeGrand:2012qa} as gauge connections.
The method of updating these gauge links is described in Ref.~\cite{DeGrand:2016pur}.
The action is written in terms of  $\beta = 2N_c / g_0^2$ and  the hopping parameter $\kappa=(2m_0^q a+8)^{-1}$
rather than the bare gauge coupling  $g_0$, the bare quark mass $m_0^q$ and the lattice spacing $a$.
The clover coefficient is fixed to its tree level value, $c_{\text{SW}}=1$.
 The gauge fields obey periodic
boundary conditions; the fermions are periodic in space and antiperiodic in time.
The lattices have $L^3\times N_t= 16^3\times 48$  sites. The lattice size and bare quark mass were chosen to minimize
finite volume effects; the product of pseudoscalar mass and spatial size $m_{PS}L$ is greater than 5.

The vector current in Eq.~\ref{eq:form} is the local current, $J_\mu= \bar \psi \gamma_\mu \psi$.
The continuum form factor is related to the lattice one by  $F(q^2)= 2\kappa Z_V F(q^2)_{latt}$
where $Z_V$ is a scheme matching factor, computed in the
``regularization independent'' or RI scheme \cite{Martinelli:1994ty}. The calculation is identical to
the one carried out for the axial vector current in Ref.~\cite{DeGrand:2023hzz}.

When needed, the lattice spacing is set via the flow 
parameter $t_0$ \cite{other,Luscher:2010iy}.
The nominal value  of the flow parameter in $N_c=3$ QCD with $N_f=2$ is
$\sqrt{t_0}=0.15$ fm according to Ref.~\cite{Sommer:2014mea}.

\subsection{Amplitudes and interpolating fields}
\begin{figure}
\centering
\begin{tikzpicture}
  \begin{feynman}
  \vertex (a) {\(I,t_0,\vec k + \vec q \)};
  \vertex [right=of a] (b);
 \vertex [below right=of b] (f1)  {\( F, \vec k\)};
 \vertex [above right=of b] (c) {\( \mu, \vec q \)};
\diagram* {
(a) -- [fermion] (b) -- [fermion] (f1),
(b) -- [boson, edge label'=\(\gamma\)] (c),
};
\end{feynman}
\end{tikzpicture}
\caption{Diagram for the pion form factor, labelling the momenta and initial and final states.
\label{fig:amp1}}
\end{figure}
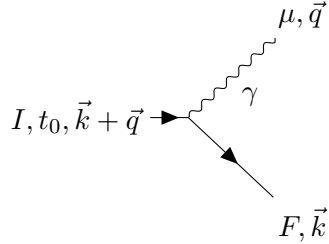

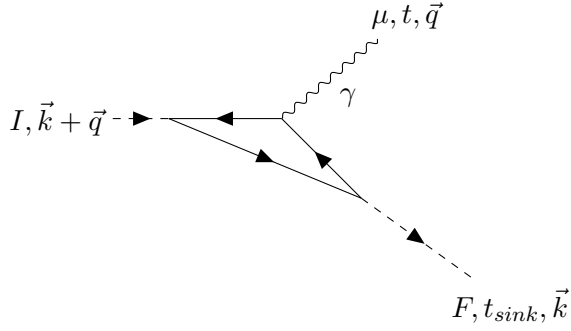
\begin{figure}
\centering
\begin{tikzpicture}
  \begin{feynman}
  \vertex (a) {\(I,\vec k + \vec q \)};
  \vertex [right=of a]  (aa)  ;
  \vertex [right=of aa] (b);
 \vertex [below right=of b] (f1) ;
 \vertex [above right=of b] (c) {\( \mu,t, \vec q \)};
\vertex [below right=of f1] (d)  {\(F, t_{sink},\vec k\)};
\diagram* {
(a) -- [charged scalar]  (aa),
(f1) -- [fermion] (b) -- [fermion]  (aa)  [blob] -- [fermion] (f1),
(b) -- [boson, edge label'=\(\gamma\)] (c),
(f1) -- [charged scalar] (d),
};
\end{feynman}
\end{tikzpicture}
\caption{Diagram at the quark level of the pion form factor, labelling the Euclidean times
of the vertices.
\label{fig:amp2}}
\end{figure}

 Graphical realizations of the necessary amplitudes are shown in 
Figs.~\ref{fig:amp1} and \ref{fig:amp2}.

The first figure works in a meson basis: a meson in an initial state $I$ with three momentum 
$\vec k+\vec q$ couples to a current emitting outgoing  momentum $\vec q$ to become
a meson in a final state $F$ carrying momentum $\vec k $.
The second figure shows the amplitude in terms of the constituent quarks and antiquarks.
As is usually done in the literature, I introduce a source interpolating field
at space-time coordinates $(\vec x_0,t_0)$, a sink or outgoing interpolating field
at time coordinate $t_{sink}$ and space coordinates $\vec y$, and put the current at
space-time coordinates $(\vec z,t)$ in between the source and sink.
I sum over $\vec y$ and $\vec z$ weighting by complex exponentials to project out
momentum eigenstates, and sew the propagators together to produce a three-point
correlator at all $t$ values between $t_0$ and $t_{sink}$. It is
 \beea
C_{FIJ_\mu}(y_0,\vec k,x_0,z_0,x_0,\vec q) &=&  \sum_{\vec y,\vec z} e^{-i\vec k \cdot \vec y}
e^{-i\vec q \cdot \vec z}
\svev{O_F(\vec y, y_0) J_\mu(\vec z,z_0) O_I(\vec x=0,\vec x)} \nonumber \\
&=&\frac{Z_F(\vec k) Z_I(\vec k+\vec q)}{4E(\vec k)E(\vec k+\vec q)} J_\mu \exp(-E(\vec k+\vec q) t)\exp(-E(\vec k)(t_{sink}-t))
\label{eq:orig3pt}
\eea
where  $y_0=t_{sink}+x_0$,  $z_0=t+x_0$, 
\bee
J_\mu=\svev{\pi(\vec k) | J_\mu(0,0) | \pi(\vec k+\vec q)}
\ee
and
\bee Z_J(\vec k)=\svev{0|O_J|\pi(\vec k)}.
\ee
The passage between the first and second lines of Eq.~\ref{eq:orig3pt}  comes by inserting complete sets of states of hadrons between the
operators and truncating  the sum to the lowest-energy state (a single pion); it further assumes that $t_{sink} \gg t \gg 0$.
The analogous formula for a two point function is
\beea
C_{FI}(y_0=t+x_0,\vec k,x_0) &=& \sum_{\vec y}e^{-i\vec k \cdot \vec y} \svev{O_F(\vec y, y_0) O_I(\vec x=0,x_0)}
\nonumber \\
&=&\frac{Z_F(\vec k) Z_I(\vec k)}{2E(\vec k)}f(E(\vec k),t) \nonumber \\
\label{eq:orig2pt}
\eea
where
\bee
f(E,t)= \exp(-Et)+\exp(-E(N_t-t)).
\ee
There are two technical  issues to face: the first is to find source and sink interpolating fields which couple to meson states with
nonzero momentum.
The second is how to extract a 
separate a measurement of $J_\mu$ from fits to a set of three point and two point functions.

At time $t_{sink}$ I use a simple point sink $\bar \psi \gamma_5 \psi$ projected on momentum $\vec k=0$.
 The fermion world line from $t_0$ to $t_{sink}$ to $t$ is computed using the usual
``sequential-source'' or ``exponentiation'' method, which is described in most texts about lattice techniques.
Being very terse, we need to compute  an object like $M= \Tr Q(x,y) T(y,x)$ 
where $Q(x,y)$ is the fermion propagator and
$T(y,x)=Q(y,z) Q(z,x)$.  Then $D(w,y)T(y,x)= Q(w,x)$, so the first propagator
is the source for the second one. 

With that construction, $C_{FIJ_\mu}$ at many values of $\vec q$ can be found without the need for additional inversions of the
Dirac operator.

Now the interpolating field at $t_0$ must couple to nonzero values of $\vec q + \vec k$. I am accustomed to gauge fixing
the link variables to Coulomb gauge and then introducing extended sources which are a product of
quark and antiquark sources $O \sim \Phi_R(\bar \psi)\Phi_R(\psi)$. 
A generalization of this construction gives a useful source.
I continue to gauge fix to Coulomb gauge.

Maybe it is clearest to introduce the interpolating fields using continuum notation.
The convention for Fourier transforms is
\bee
\psi(\vec p)= \int d^3xe^{i\vec p \cdot \vec x}\psi(\vec x); \qquad 
\psi(\vec x)= \int \frac{d^3p}{(2\pi)^3} e^{-i\vec p \cdot \vec x}\psi(\vec p). 
\ee
Then a Gaussian source centered at the origin is
\beea
\Phi_R(\psi) &=& \int d^3x S(x) \psi(\vec x) \nonumber \\
&=& \int d^3x e^{-x^2/R^2} \psi(\vec x)  \nonumber \\
&=& C \int \frac{d^3p}{(2\pi)^3}e^{-p^2R^2/4}\psi(\vec p) \nonumber \\
&=& \int d^3 p S_R(\vec p)\psi(p)
\eea
and we see that its strength as a creation operator of  a particle of momentum $\vec p$
dies as $\vec p$ grows.

The solution of Ref.~\cite{Izubuchi:2019lyk} is to replace the Gaussian source by
something which obviously peaks at $\vec p= \vec K$:
\bee
S_{\vec K}(\vec p) \propto \exp(-\frac{R^2}{4}(\vec p - \vec K)^2 )
\ee
from which I define
\bee
S_{\vec K}(\vec x)= N
\int \frac{d^3p}{(2\pi)^3} e^{-i\vec p \cdot \vec x}  \exp(-\frac{R^2}{4}(\vec p - \vec K)^2).
\label{eq:sk}
\ee
(The normalization will only be needed when $Z$ factors are quoted below.)
The authors of Ref.~\cite{Izubuchi:2019lyk} work directly with a meson interpolating field
\beea
\Phi_{\vec K_1,\vec K_2}(x)  &=& \int d^3 x'' d^3 x' \bar \psi(x'')
 S_{\vec K_1}(x''-x)\Gamma S_{\vec K_2}(x-x') \psi(x') \nonumber \\
&=& \int  \frac{d^3q_1}{(2\pi)^3} \frac{d^3q_2}{(2\pi)^3} \bar \psi(q_1)\Gamma \psi(q_2)
\exp[-\frac{R^2}{4}\left( (q_1+K_1)^2 + (q_2 - K_2)^2 \right) ] \nonumber \\
\eea
which peaks at $q_1= - K_1$ and $q_2 = K_2$.  And
\beea
\Phi_{\vec Q} &=& \int d^3x e^{i\vec Q\cdot \vec x} \Phi_{\vec K_1,\vec K_2}(x) \nonumber\\
&=& \int \frac{d^3q_1 d^3q_2}{(2\pi)^3}\delta^3(\vec Q - \vec q_1 - \vec q_2)
 \bar \psi(q_1)\Gamma \psi(q_2)
\exp[-\frac{R^2}{4}\left( (q_1+K_1)^2 + (q_2 - K_2)^2 \right) ] \nonumber \\
\eea
projects out a $\vec Q$ momentum eigenstate. Clearly the coupling strength
is greatest when $\vec Q= \vec K_2 - \vec K_1$, and this suggests taking $\vec K_2=\vec Q/2$,
$\vec K_1= -\vec Q/2$ as an optimal choice for a source.

The correlator of an extended source and extended sink also factorizes,
\beea
C_{SS}(y_0, \vec p, \vec x) &=& \sum_{\vec y} \exp(-i\vec p\cdot\vec y)
\svev{\Phi_{-\vec K,\vec K}(y,y_0) \Phi_{ -\vec K,\vec K}(x,x_0)} \nonumber \\
&=& \sum_{\vec y,\vec y', \vec y'', \vec x', \vec x''} \exp(-i\vec p\cdot\vec y)   \times \nonumber \\
& & \Tr(
[S_{-\vec K}(x-x'') Q(x'',y'') S_{-\vec K}(y''-y)]
\Gamma
[S_{\vec K}(y-y') Q(y',x') S_{\vec K}(x'-x)]
\Gamma
) \nonumber\\
\eea
(with $\Gamma=\gamma_5$ here; clearly the procedure generalizes to any meson interpolating field).
 Of course
\bee
D_{zy'}(Q(y',x) S_{\vec K}(x-x') )  =      S_{\vec K}(z-x') 
\ee
that is, one performs the inversion on the smeared source. 
Then the full correlator is assembled using the ``meta-propagators''
\bee
{\cal Q}_{\vec K}(y,x)=\sum_{\vec y', \vec x'} S_{\vec K}(y-y')Q(y',x') S_{\vec K}(x'-x).
\ee
Notice that a separate inversion must be done for the quark line and the antiquark line because the
sources involve $-\vec K$ and $\vec K$.

Passage to the lattice involves only a few small changes. The momentum integrals in the formulas above become
sums over discrete values defined across the Brillouin zone. The weighting momentum $K$, however, can take any value.
To prevent discontinuities, 
$\vec p- \vec K$ is defined to be the shortest distance 
between $\vec p$ and $\vec K$ in the Brillouin zone.
All convolutions are performed using fast Fourier transforms.

I experimented with various choices for $\vec K$ by looking at spectroscopy on $16^3\times 32$ lattices.
Pure Gaussian sources ($\vec K=0$ in the above description) could barely see the lowest nonzero momentum
mesons. A choice of $\vec K=(0.2,0.2,0.2)$ was able to capture a signal for $\vec q=(1,0,0)$, $(1,1,0)$ and $1,1,1)$ times the fundamental
scale  of $2\pi/16$. 
A $q$ value of $2\pi/16$ is 0.4 and the optimum $K$ from the analysis above is $K=q/2$ which is not too different from my choice.
The source could see the $\vec q=(2,0,0)2\pi/L$ state, though this was very noisy.
 I could not find a $\vec K$ value
which could produce a high quality signal for it: that is a problem with small lattices, even the minimum nonzero momentum is not small.
I did not try very hard to do better, however.

I kept $R=6$ in the definition of $S_{\vec K}(\vec p)$ since that worked well for 
zero momentum spectroscopy with the present set of bare parameters.

So to proceed: I have one final state momentum, $\vec k=0$. 
$t_{sink}$ is fixed to the value 22, almost halfway across the $N_t=48 $ lattice.
I analyzed  the three-point  correlator for $t$ between zero and $t_{sink}$. I recorded correlators for
eleven different values of $\vec q$: in units of $2\pi/L$, $\vec q=(0,0,0)$, $(1,0,0)$, $(0,1,0)$, $(0,0,1)$, then
$(1,1,0)$ plus permutations,  then $(2,0,0)$ plus permutations, and $(\vec q=1,1,1)$. I averaged the correlators over all
lattice -symmetric points. For matrix elements of $J_0$, this is an average over all permutations of
positive $\vec q$ values. (Note $\vec K$ in the source is not symmetric under reflection.)  Since
$\svev{\pi(\vec k=0) | J_i(q) | \pi(\vec q)} = q_i F(q^2)$, I combined measurements of $J_i$ with corresponding $q_i$
values which were expected to give a common value of $F$: for example, $J_1$ with $\vec q=(1,0,0)$, $J_2$ with $\vec q=(0,1,0)$,
and $J_3$ with $\vec q=(0,0,1)$. ``Odd'' combinations, such as the correlator for $J_1$ computed at $\vec q=(0,1,0)$ averaged  to zero, noisily.

I should note (and this is explained nicely by Ref.~\cite{Izubuchi:2019lyk}) -- the relevant part of the correlator for
matrix elements of the timelike current $J_0$ is its real part, while the relevant part of the
correlator for spacelike $J_i$ is the imaginary part.

\subsection{Fitting strategy}

Let me again collect all the fitting functions needed for the actual analysis.
There are two three-point correlators,
\bee
C_\mu(t,t_{sink}) = \frac{Z_{SI}(\vec p_I) Z_{PF}(\vec k)}{4E_I E_F} V_\mu \exp(-E_I t)\exp(-E_F(t_{sink}-t))
\label{eq:cmu}
\ee
where $V_\mu=\svev{\pi(\vec p_F) | J_\mu | \pi(\vec p_I)}$
and
\bee
C_F(t,t_{sink}) = \frac{Z_{SI}(\vec p_I) Z_{PF}(\vec k)}{4E_I E_F} (E_I+E_F)F \exp(-E_I t)\exp(-E_F(t_{sink}-t))
\label{eq:cf}
\ee
where $F=F(q^2)_{latt}$. (I am being a bit redundant with my notation; $I$ and $F$ label the initial
and final
state meson, and, once more, $\vec p_f=\vec k$ and $\vec p_I=\vec k+ \vec q$.)

For each particle momentum $(J=I,F)$ there are two kinds
of two point functions:  a smeared source and a smeared sink,
\bee
C_{SS}(\vec p_J,t)=\frac{Z_{S}(\vec p_J)^2}{2E_J}f(E_J,t);
\label{eq:cssj}
\ee
and a smeared source and a point sink
\bee
C_{PS}(\vec p_J,t)=\frac{Z_{S} (\vec p_J)Z_{P}(\vec p_J)}{2E_J}f(E_J,t).
\label{eq:cpsj}
\ee

$V_i$ and $F$ are the interesting quantities. The literature describes  many ways to extract them.
The most common one is to take ratios of three point and two point functions in which the $Z$'s cancel.
The problem with this, in my opinion, is that many of the methods involve fractional powers of 
the two point functions (for an example, see Ref.~\cite{QCDSFUKQCD:2006gmg}), and an issue is that the two point functions 
can fluctuate to negative values.
It seemed to me that it was easier and potentially more transparent to perform correlated fits to a three point functions and a collection of two point functions.
So that is what I did.  In doing this I was motivated by the analysis in Ref.~\cite{FermilabLattice:2022gku}.

So a calculation of $F$ or $V_\mu$ requires a simultaneous fit to four
two point correlators, $C_{SS}(\vec p_I)$, $C_{PS}(\vec p_I)$, $C_{SS}(\vec p_F)$ and $C_{PS}(\vec p_F)$ and to one of the three point
correlators, Eqs.~\ref{eq:cmu} or \ref{eq:cf}. The fit has seven free parameters,
$V_\mu$ or $F$ plus $Z_{SS}(\vec p_I)$, $Z_{SP}(\vec p_I)$, $E_I$, $Z_{SS}(\vec p_F)$, $Z_{SP}(\vec p_F)$ and $E_F$.

The $\vec q=0$ form factor can be found more simply, with a simultaneous fit to the two point
function
\bee
C_{PS0}(t)= A_0f(m_{PS},t)
\label{eq:sp0}
\ee
and the three point function
\bee
C_0(t,t_{sink}) = \frac{A_0}{2m_{PS}} (2m_{PS})F(0)\exp(-m_{PS}t_{sink}).
\label{eq:c0}
\ee
(I've kept the redundant factors for comparison with Eq.~\ref{eq:cf}.)
There are three parameters to be fit, $A_0$, $m_{PS}$ and $F(0)=F(q^2=0)_{latt}$.

 All results come from a standard full correlated
analysis involving fits to a wide range of $t$'s.
Best fits are chosen with the ``model averaging'' ansatz of Jay and Neil \cite{Jay:2020jkz}.
Recall that this method assigns each particular fit in a suite (with a chi-squared value $\chi^2$ and
$N_{DOF}$ degrees of freedom) a weight in the average which is proportional to
$\exp(-(\chi^2/2 - N_{DOF}))$.
Typically, I performed a fit for a particular $\vec q$ parameter value as follows: I first did a  three parameter combined fit to $C_{SS}$ and $C_{SP}$
and model-averaged the result to find a range if $t$ values for the two correlators which had high weights in the model average.
I then took sets of these $t$ values and did  five-correlator fits whose output was either $F$ or $V$ (plus everything else).
I  first performed ``window'' fits where I walked across the $t$ values kept in Eqs.~\ref{eq:cmu} - \ref{eq:cf}, with $t_{max}=t_{min}+4$.
I found a range where the fit values of $F$ or $V_\mu$ appeared flat to the eye. I then resumed doing fits slightly outside this range,
varying $t_{min}$ and $t_{max}$ independently, and passed the results
 through the model averaging filter.
With model averaging it is possible to do a large number of fits and the averaging will just discard (or give a tiny weight) to
 fits which have high chi-squared.
Therefore,
what I did might not be acceptable to the over cautious reader, but I actually wanted to look
 at the fits myself rather than
just take what came out of the model averaging black box.

\section{Results\label{sec:results}}
\begin{figure}
\begin{center}
\includegraphics[width=0.8\textwidth,clip]{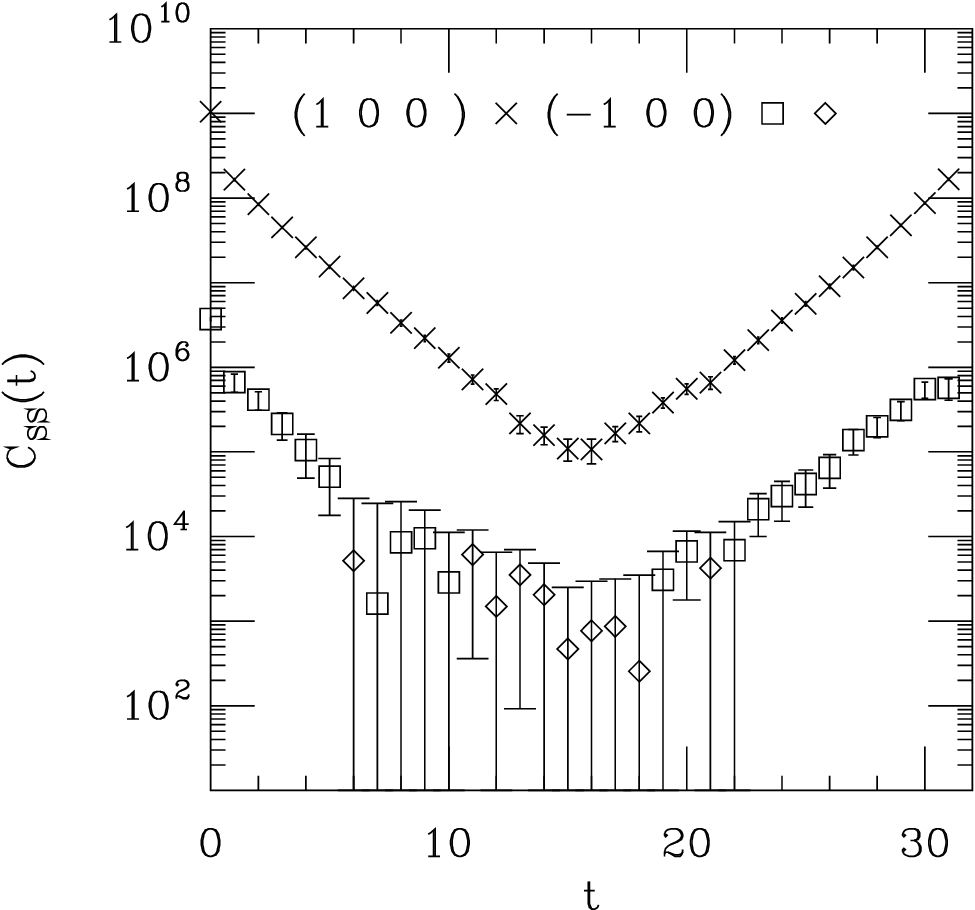}
\end{center}
\caption{ Correlators $C_{SS}(\vec q)$ for a favorable combination of $\vec q$ and $\vec K$
(crosses) and an unfavorable combination (squares, and diamonds when the correlator is negative),
from an $SU(3)$ data set at $\beta=5.4$, $\kappa=0.127$ on volume $16^3\times 32$.
\label{fig:wps12}}
\end{figure}

I will start with some preliminary pictures. Fig.~\ref{fig:wps12} shows 
$C_{SS}(\vec q)$  for $q=(\pm1,0,0)$ from an $SU(3)$ data set with identical  bare parameters
to the ones used for the form factor, but lattice 
volume $16^3\times 32$ sites. The source had $\vec K = (0.2,0.2, 0.2)$ which is
a favorable source for $\vec q=(1,0,0)$ but an unfavorable source for $\vec q=(-1,0,0)$
The correlator for the unfavorable combination of $\vec q$ and $\vec K$ is clearly smaller and noisier.
(It even has negative entries, shown as diamonds in the figure.)

\begin{figure}
\begin{center}
\includegraphics[width=0.8\textwidth,clip]{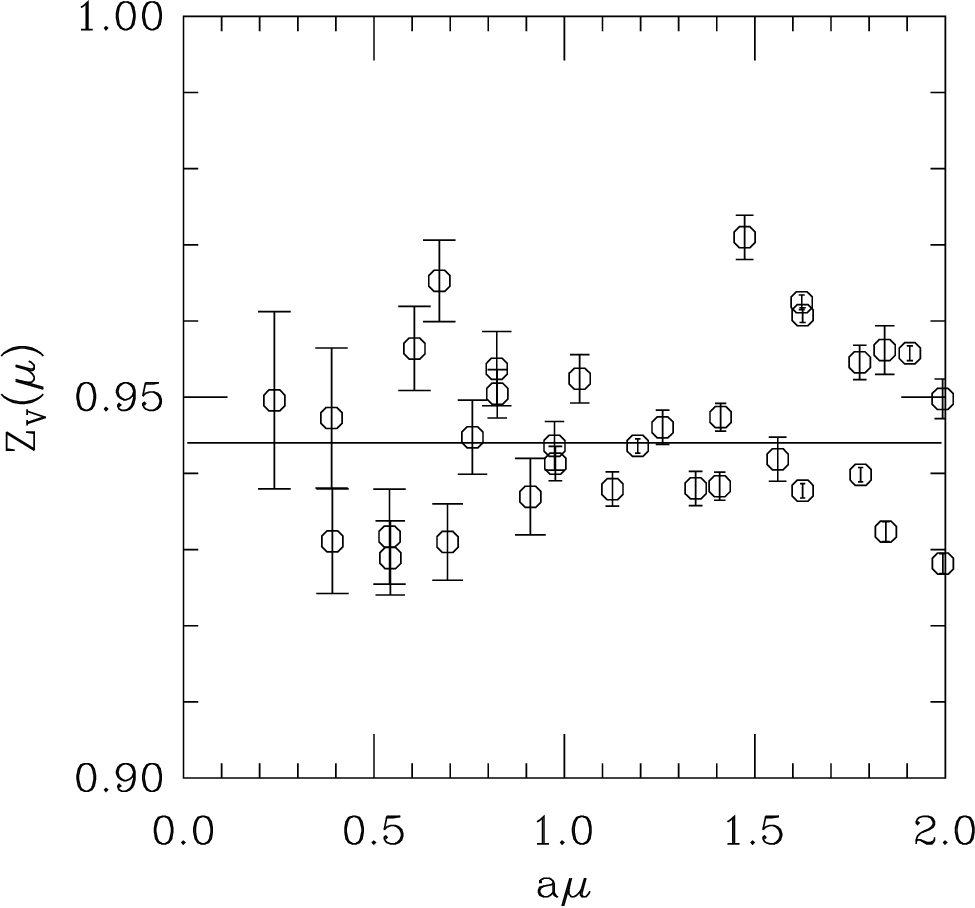}
\end{center}
\caption{ $Z_V(\mu)$ versus $a\mu$ from $16^3\times 32$ $SU(4)$ data sets at
 $\beta=10.2$, $\kappa=0.1265$.
\label{fig:vvsu4}}
\end{figure}

As remarked earlier, the lattice to continuum current renormalization factor $Z_V$ is computed 
 in the
``regularization independent'' or RI scheme \cite{Martinelli:1994ty}.
I used pre-existing $16^3\times 32$ (for $SU(4)$ and $SU(5)$) and $24^3\times 32$ (for $SU(3)$)
data sets to measure $Z_V$. Full details can be found in Ref.~\cite{DeGrand:2023hzz}.
$Z_V(\mu)$ is expected to be independent of the scale $a\mu$ where it is evaluated.
Because the spatial size of the lattices used is small, lattice artifacts are visible in a plot
of $Z_V$ versus $a\mu$ where $(a\mu)^2$ is the squared magnitude of the four momentum at which
the measurement is performed ($(a\mu)^2=\sum_j a^2 p_j^2$). Fitting data from small regions
 of $\mu$ produces an uncertainty
on the order of 0.002 to $Z_V$ (whose mean value is close to unity) but such fits have
poor confidence level due to the lattice artifact scatter. A figure  (the $SU(4)$ case
is shown in Fig.~\ref{fig:vvsu4}) shows that the scatter is about a per cent at $a\mu=1.0$
or $\mu\sim$ 2 GeV, which is where I choose to  evaluate $Z_V$. I will take that as an uncertainty for $Z_V$.
 We will shortly see that the
uncertainty on $F(q^2)_{latt}$ is always greater than a few per cent (sometimes much greater)
and so this  choice is not going to affect the final results.
The three values of $Z_V$ are
0.94(1)  for $SU(3)$,
0.94(1)  for $SU(4)$,
and
0.95(1)  for $SU(5)$.

\begin{figure}
\begin{center}
\includegraphics[width=0.8\textwidth,clip]{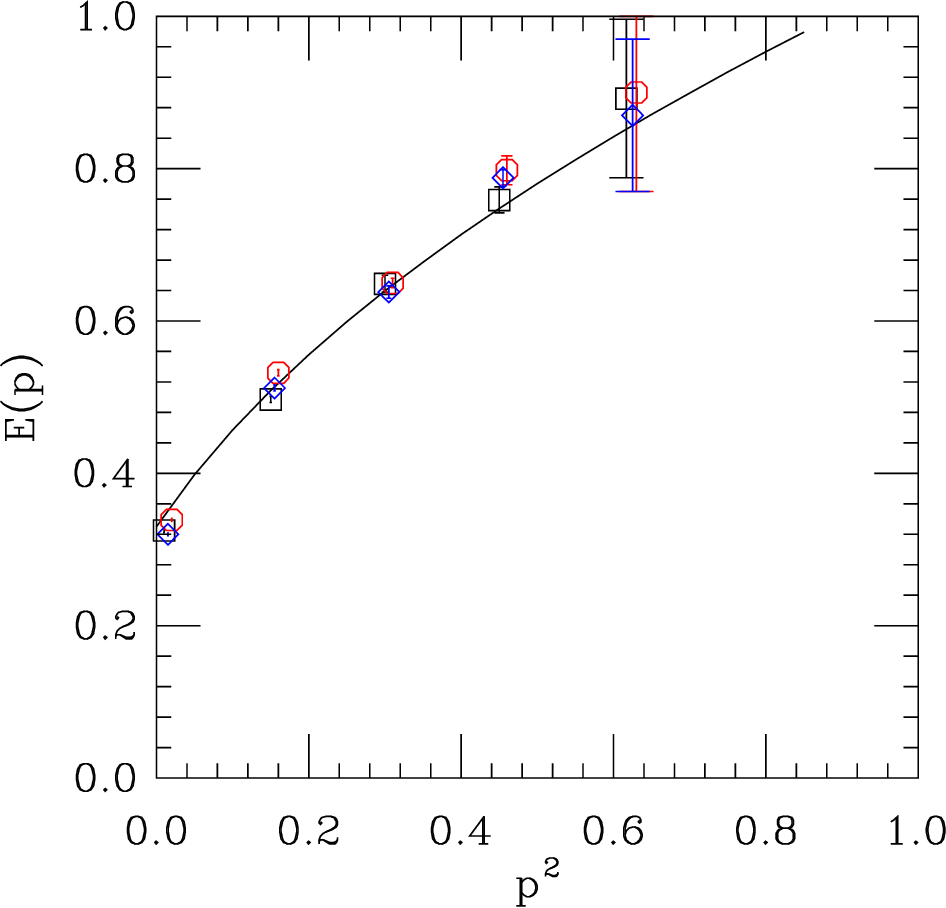}
\end{center}
\caption{ Energy as a function of momentum (all in lattice units)  from fits to simulation data.
 The line is the
expected continuum result $E=\sqrt{p^2+m^2}$ with $m=0.33$, roughly the
common pion mass of all three colors (compare Tab.~\protect{\ref{tab:params}}).
\label{fig:energy}}
\end{figure}

Next comes the dispersion relation $E(\vec p)$ versus momentum. When the three momentum gets large,
one expects to see a lattice dispersion relation, not a continuum one.
Fig.~\ref{fig:energy} shows the energy  in lattice units, $E(\vec p)$ as a function of $p^2$, 
for the different $N_c$ values
and $p$ values. The curve is a plot of $E(\vec p)=\sqrt{m^2+p^2}$  versus $p^2$ with $m=0.33$, the common 
pion mass for all colors. The data seem to be consistent with a continuum dispersion relation.
The $p=(2,0,0)$ or $p^2=(\pi/4)^2$ energies are quite noisy; more on these data set below.

Now for fits of $F$ or $V_x$. The overwhelming impression as one moves from one $N_c$ value
to another is how nearly identical the data sets are.

\begin{figure}
\begin{center}
\includegraphics[width=0.8\textwidth,clip]{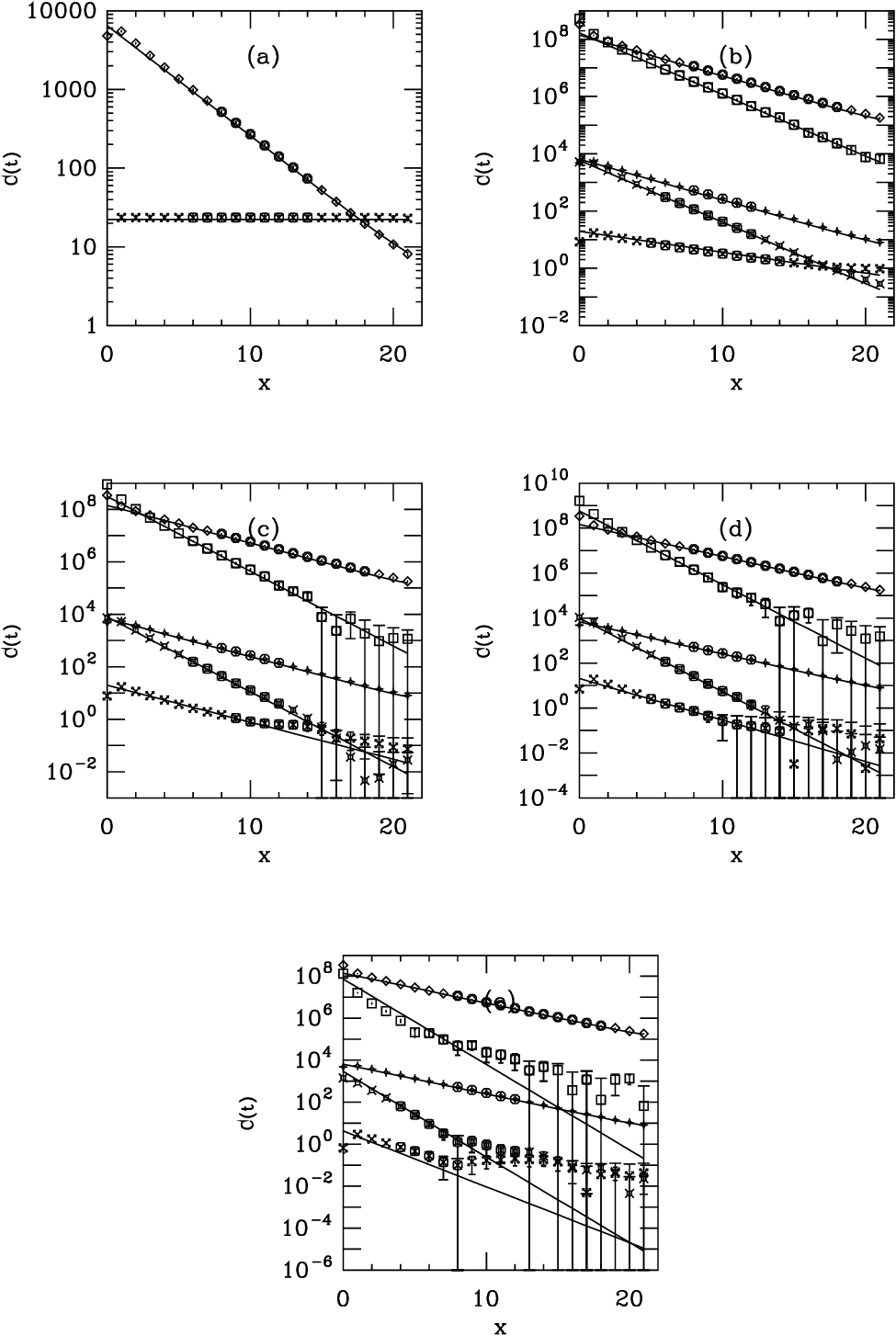}
\end{center}
\caption{Examples of fits producing $F(q^2)$ for the $SU(3)$ data set. Panels are
(a) $q=(0,0,0)$;
(b) $q=(1,0,0)$;
(c) $q=(1,1,0)$;
(d) $q=(1,1,1)$;
(e) $q=(2,0,0)$ (all in units otf $2\pi/L$). In all figures, the points used in the fits are marked as octagons.
Lines are fits to the individual correlators. In all cases the three point correlator is 
labelled by a fancy cross. In panel (a), diamonds label the two point function. In the other
figures,
squares and fancy squares label $C_{SS}(t)$ and $C_{SP}(t)$ for the initial momentum ($\vec p = \vec k+\vec q$)
while 
 $C_{SS}(t)$ and $C_{SP}(t)$ for the final momentum ($\vec p = \vec k$) are labelled by
diamonds and fancy diamonds.
\label{fig:allft0su3}}
\end{figure}

\begin{figure}
\begin{center}
\includegraphics[width=0.8\textwidth,clip]{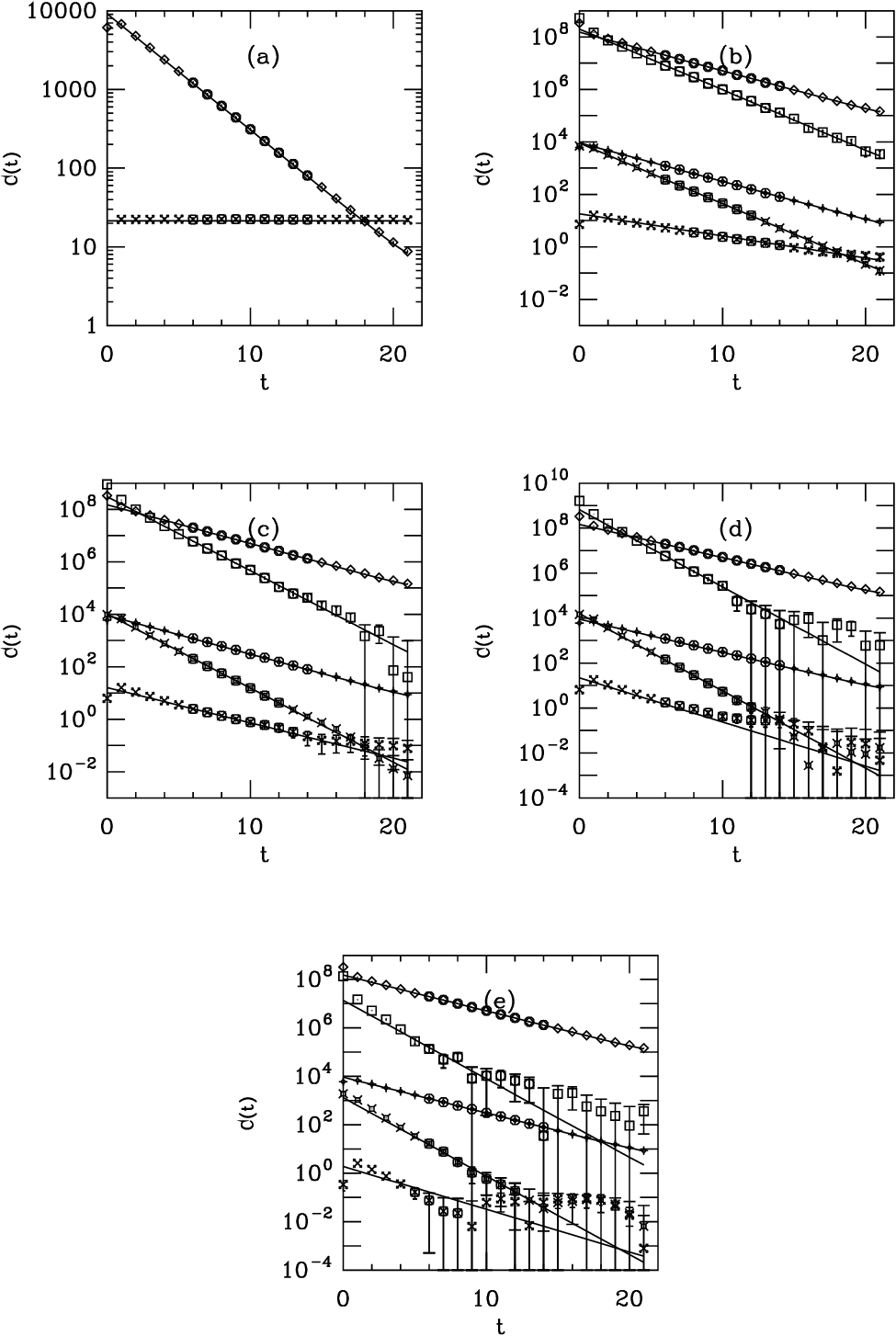}
\end{center}
\caption{Examples of fits producing $F(q^2)$ for the $SU(4)$ data set.
The panels and presentation are the same as for the $SU(3)$ set, Fig.~\protect{\ref{fig:allft0su3}}.
\label{fig:allft0su4}}
\end{figure}

\begin{figure}
\begin{center}
\includegraphics[width=0.8\textwidth,clip]{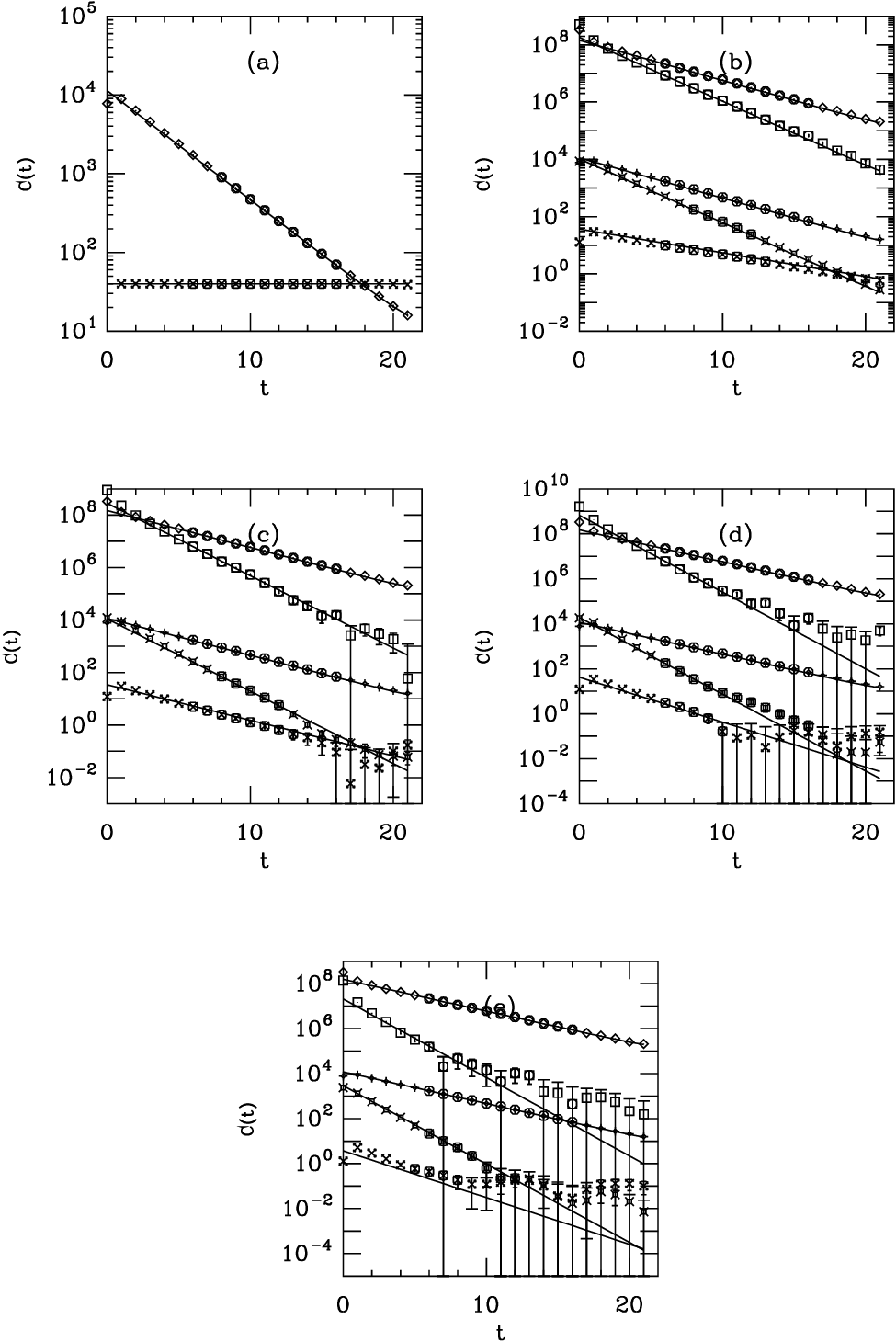}
\end{center}
\caption{Examples of fits producing $F(q^2)$ for the $SU(5)$ data set.
The panels and presentation are the same as for the $SU(3)$ set, Fig.~\protect{\ref{fig:allft0su3}}.
\label{fig:allft0su5}}
\end{figure}

Three figures, one for each $N_c$, illustrate this point, plots of fits to 
$C_{PS0}(t)$ and $C_0(t,t_{sink})$ to Eqs.~\ref{eq:sp0} and \ref{eq:c0}
(which give $F(q^2=0)$) and to $C_{SS}(\vec p)$ and $C_{PS}(\vec p)$
 (for $\vec p= \vec k$ and $\vec p=\vec k+\vec q$) to Eqs.~\ref{eq:cssj} and \ref{eq:cpsj}
along with $C_F(t,t_{sink})$ to Eq.~\ref{eq:cf}.
These are shown in Figs.~\ref{fig:allft0su3}, \ref{fig:allft0su4} and \ref{fig:allft0su5}.

\begin{figure}
\begin{center}
\includegraphics[width=0.8\textwidth,clip]{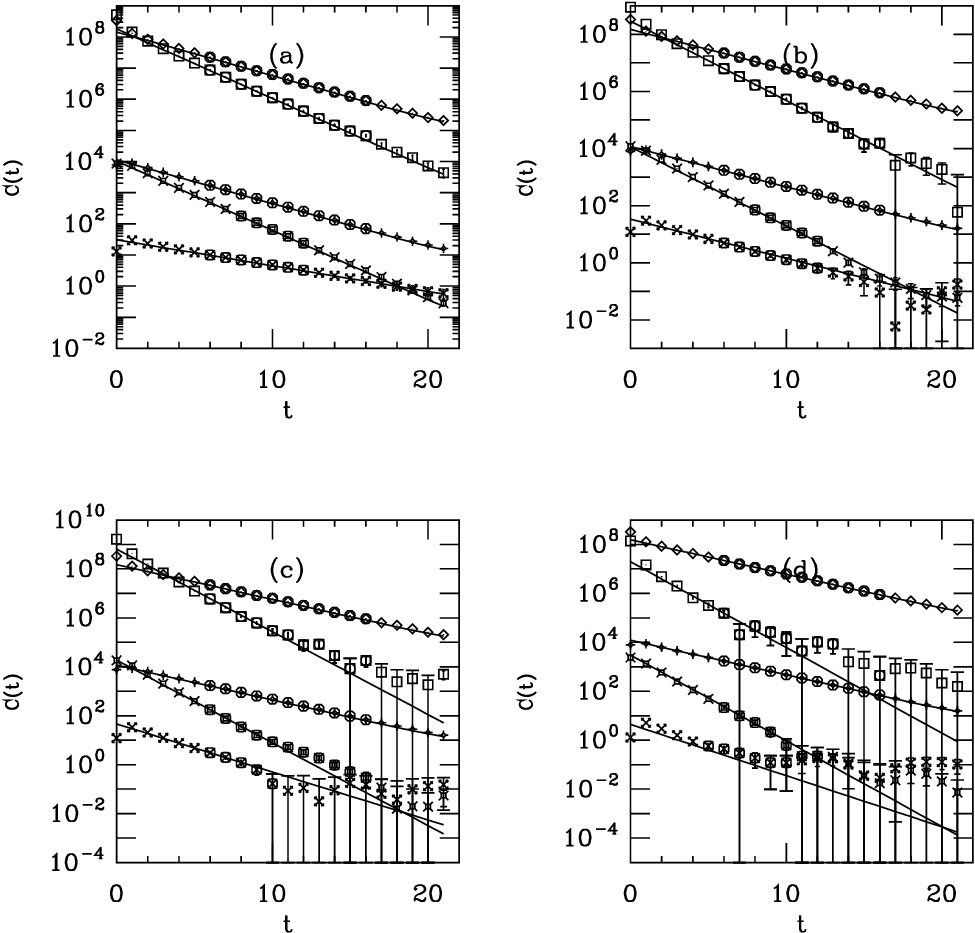}
\end{center}
\caption{Examples of fits producing $V_x$ for the $SU(5)$ data set.
 Panels are
(a) $q=(1,0,0)$;
(b) $q=(1,1,0)$;
(c) $q=(1,1,1)$;
(d) $q=(2,0,0)$ (in units of $2\pi/L$).
The presentation is the same as for the figures showing fits producing $F(q^2)$.
\label{fig:all3xsu5}}
\end{figure}

For completeness I will show one fit to $V_x$, for the $SU(5)$ date set, in Fig.~\ref{fig:all3xsu5}.

There are many consistency checks of the fits. The final momentum is $\vec k=0$ and so
the associated $Z$ factors and pseudscalar masses should be identical across all $\vec q$ values.
$A_0$ from Eq.~\ref{eq:sp0} should match $Z_P Z_S/(2m_{PS})$ from the nonzero $\vec q$ fits.
The initial state $Z$ factors and energies should match between $F$ and $V_x$ fits.
All these checks produce consistent results.
Table \ref{tab:nuisance} gives results for the  $Z$'s and $E$'s. (The overall scale of $S_{\vec K}(x)$
is not physical (in Eq.~\ref{eq:sk} $N=R^3\pi^{3/2}$)
 but as long as it is the same for all $\vec p$ values
one can use the relative sizes of $Z_S$'s to compare how well the
 source couples to different states.)
 The $Z$'s show rough $\sqrt{N_c}$ scaling
as expected. They also show the efficiency of the coupling of the source to each momentum eigenstate:
the largest coupling is to the $\vec p = (1,1,1)$ state, as is also expected.

And I can make a few remarks on the data at different $q$ values. The case $\vec q=(2,0,0)$ is 
obviously the noisiest one,
with barely a signal to report. 
One issue is the source, which simply did not couple well to
this $\pi(\vec q)$ state. I spent a little time experimenting with alternative sources
(after all, the choice of $K$ values for the source was empirical) but I did not find
a better choice. I think this is just an issue with using a small spatial lattice:
$q=2(2\pi/16) =0.79$ is already a large number.
This momentum state also has the largest energy in the ensemble 
and the  Parisi \cite{Parisi:1983ae} -  Lepage \cite{Lepage:1989hd} argument
says that the signal to noise ratio, proportional to $\exp((m_{PS}-E(\vec p))t)$, is the worst in the data set.

The other $\vec q$ values are better behaved. As an example, the ranges of $t$ values with higher weights
in fits to $C_F(t,t_{sink})$ for $N_c=5$ were
(5-7) to (13-17) for $\vec q= (1,0,0)$,
(8-10) to (12-16) for $\vec q= (1,1,0)$,
(5-7) to (12-16) for $\vec q= (1,1,1)$,
but only (4-5) to (7-9) for $\vec q = (2,0,0)$.

\begin{figure}
\begin{center}
\includegraphics[width=0.8\textwidth,clip]{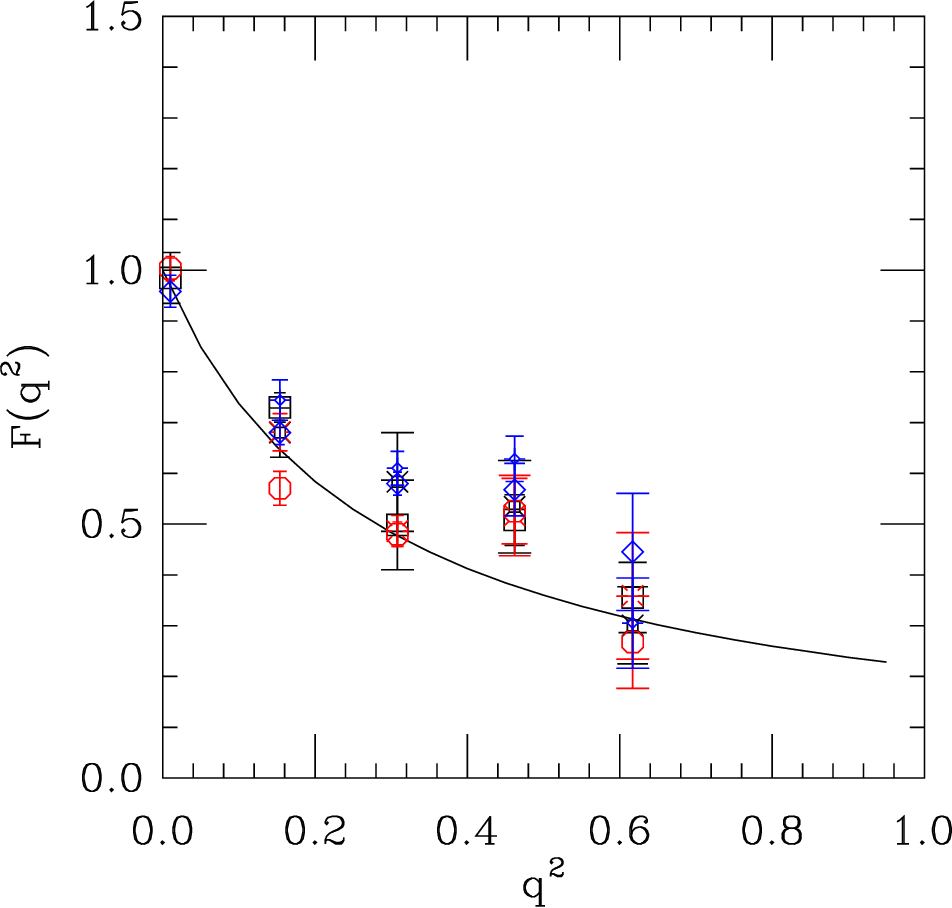}
\end{center}
\caption{ The  form factor of the pseudoscalar meson as a function of squared momentum 
transfer in lattice units.
Squares label $SU(3)$ points from the $\mu=0$ current; fancy squares label points from $J_x$.
Red octagons and bursts label the corresponding points for $SU(4)$
and blue diamonds and fancy diamonds repeat the assignment for $SU(5)$.
The line is the expectation from the vector meson dominance, Eq.~\protect{\ref{eq:vdm}}, using a vector meson mass of
$am_V=0.53$.
\label{fig:form}}
\end{figure}

Finally, the data for the form factor is shown in Fig.~\ref{fig:form} and is recorded in Tables \ref{tab:form0} and \ref{tab:formx}.
The shape of $F(q^2)$ is independent of the number of colors.

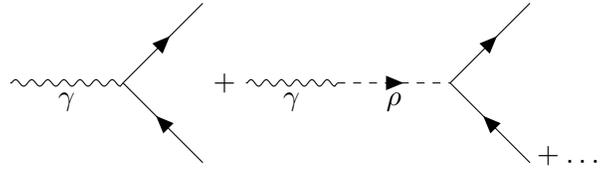
\begin{figure}
\centering
\begin{equation}
\begin{tikzpicture}
  \begin{feynman}
  \vertex (a);
  \vertex [right=of a] (b);
 \vertex [below right=of b] (f1) ;
 \vertex [above right=of b] (c);
\diagram* {
(f1) -- [fermion] (b) -- [fermion] (c),
(a) -- [boson, edge label'=\(\gamma\)] (b),
};
\end{feynman}
\end{tikzpicture}
\begin{tikzpicture}
  \begin{feynman}
  \vertex (aa) {\(+\)};
\vertex [right=of aa] (a);
  \vertex [right=of a] (b);
  \vertex [right=of a] (b);
 \vertex [below right=of b] (f1) ;
 \vertex [above right=of b] (c);
\diagram* {
(f1) -- [fermion] (b) [dot]-- [fermion] (c),
(aa) -- [boson, edge label'=\( \gamma\)] (a), 
(a) -- [charged scalar, edge label'=\( \rho\)] (b),
};
\end{feynman}
\end{tikzpicture}
+\dots
\nonumber
\end{equation}
\caption{Amplitudes which contribute to the vector dominance formula, Eq.~\protect{\ref{eq:vdm2}}. The two
diagrams correspond to the two
terms in the right hand side of this equation.
\label{fig:vdm}}
\end{figure}

The line in Fig.~\ref{fig:form} is not a fit, it is the simplest possible parameterization from
the vector dominance model (\cite{vdm}; see also \cite{Klingl:1996by})
\bee
F(q^2) = \frac{1}{1+\frac{q^2}{m_V^2}}
\label{eq:vdm}
\ee
where $am_V=0.53$, the common vector meson mass for the data sets as shown in 
Table \ref{tab:params}.
The model comes from considering diagrams like the ones in 
Fig.~\ref{fig:vdm}.
Klingl et al \cite{Klingl:1996by} present several variations on this formula,
using
the definition of the rho to vector meson coupling as $\svev{\gamma|\rho}=e m_V^2/g_V$
and the direct coupling of a vector to two pseudoscalars of 
\bee
{\cal L} = \frac{ig_{VPP}}{4}\Tr(V_\mu[\partial_\mu \pi, \pi] )
\ee
which gives
\bee
F_{VDM}(q^2) =  1 - \frac{g_{VPP}}{g_V}\frac{q^2}{q^2+m_V^2+\Pi(q^2)}
\label{eq:vdm2}
\ee
and $\Pi(q^2)$ is the vector meson vacuum polarization, a two-pion loop.
We already know that $1/g_V\propto\sqrt{N_c}$ (see, for example,  Ref.~\cite{DeGrand:2016pur})
and the $N_c$ independence of $F(q^2)$ tells us that $g_{VPP}$ should scale as $1/\sqrt{N_c}$,
or that the hadronic decay width of the vector meson should scale as $1/N_c$, which
is precisely what large $N_c$ counting predicts. There will be $1/N_c$ corrections to the form factor
from $\Pi(q^2)$ which could remain a target for a better study if desired.

\section{Summary\label{sec:summary}}
Once again, a low statistics calculation of an observable across $N_c$ shows agreement with expectations based on
large $N_c$ counting. In the language of the quark model, the charge density of the pseudoscalar meson is independent of the
color group. Stepping back a bit, from the point of view of numerical simulation, there
  seems to be nothing special about QCD;  $SU(N_c)$ gauge theories  with a small
number of flavors of dynamical fundamental fermions show similar behavior and the differences are well described by
large $N_c$ counting rules. 
\clearpage

\begin{table}
\begin{tabular}{c c c c}
\hline
    & $SU(3)$ & $SU(4)$ & $SU(5)$ \\
\hline
$\beta$   & 5.4 & 10.2 & 16.4 \\
$\kappa$  & 0.127 & 0.1265 & 0.1265 \\
$t_0/a^2$ & 2.155(7) & 2.312(10) & 2.386(6) \\
$am_{PS}$  & 0.328(1) & 0.341(2) & 0.323(1) \\
$am_V$    & 0.531(5) & 0.529(3) & 0.534(3) \\
\hline
\end{tabular}
\caption{Relevant lattice quantities for the project. The flow parameter $t_0/a^2$ and the pseudoscalar mass
in lattice units $am_{PS}$ are taken from Ref.~\cite{DeGrand:2023hzz}
and the vector meson mass in lattice units $am_V$ is taken from Ref.~\cite{DeGrand:2016pur}.
\label{tab:params}}
\end{table}

\begin{table}
\begin{tabular}{c  | c c c c}
\hline
 $N_c$ & $p$ & $Z_S$ & $Z_P$ & $E(p)$ \\
 \hline
 3 & $(0,0,0)$  &  $9.5(1)\times 10^3$    &   0.45(1)    &   0.325(3)  \\
    & $(1,0,0) $ & $1.29(2)\times 10^4$     & 0.46(1)     &   0.497(4)\\
    & $(1,1,0) $ & $1.95(8)\times 10^4$     & 0.50(2)     &   0.649(11)\\
    & $(1,1,1) $ & $3.0(2)\times 10^4$     & 0.53(4)     &   0.759(17)\\
    & $(2,0,0) $ & $1.0(3)\times 10^4$     & 0.4(1)     &   0.89(10)\\
    \hline
 4 & $(0,0,0)$  & $1.00(1)\times 10^4$     &   0.614(7)     &     0.339(2)  \\
    & $(1,0,0) $ & $1.45(2)\times 10^4$     &  0.66(1)    &     0.532(4) \\
    & $(1,1,0) $ &  $1.99(4)\times 10^4$    &  0.66(1)    &    0.650(6) \\
    & $(1,1,1) $ &  $3.30(2)\times 10^4$    &  0.79(4)    &    0.798(19) \\
    & $(2,0,0) $ &  $5.2(12)\times 10^3$    &  0.48(15)    &   0.90(13) \\
\hline
 5 & $(0,0,0)$  &  $9.75(8)\times 10^3$    &   0.760(6)    &      0.320(3) \\
    & $(1,0,0) $ & $1.33(3)\times 10^4$     &  0.77(2)    &   0.512(3) \\
    & $(1,1,0) $ & $1.91(6)\times 10^4$     &  0.76(3)    &   0.638(3) \\
    & $(1,1,1) $ & $3.2(1)\times 10^4$     &  0.94(4)    &    0.788(3)\\
    & $(2,0,0) $ & $6.8(14)\times 10^3$     &   1.0(2)   &   0.87(10) \\
\hline
 \end{tabular}\caption{``Nuisance parameters'' in the fits: the two vertex factors for two point functions and the energy of the propagating
state. The momentum $\vec p$ is given in units of $2\pi/L$ or $\pi/8$.
\label{tab:nuisance}}
\end{table}

\begin{table}
\begin{tabular}{c | c c c}
\hline
 $N_c$ & $q$ &  $F(q^2)_{latt}$  & $F(q^2)$ \\
 3 & $(0,0,0)$ &   4.13(21)  &  0.99(5)    \\
   & $(1,0,0)$ &   3.06(12)  &  0.73(3)   \\
   & $(1,1,0)$ &   2.09(37)  &  0.51(9)  \\
   & $(1,1,1)$ &   2.13(21)  &  0.51(5)  \\
   & $(2,0,0)$ &   1.49(29)  &  0.36(7)  \\
    \hline
 4 & $(0,0,0)$ &   4.20(10)  &  1.00(3) \\
   & $(1,0,0)$ &   2.39(14)  &  0.57(3)   \\
   & $(1,1,0)$ &   2.01(10)  &  0.48(2)  \\
   & $(1,1,1)$ &   2.20(27)  &  0.52(6)   \\
   & $(2,0,0)$ &   1.12(38)  &  0.27(9)   \\
\hline
 5 & $(0,0,0)$ &   4.00(10)  &  0.96(3)  \\
   & $(1,0,0)$ &   2.84(8)   &  0.68(2)  \\
   & $(1,1,0)$ &   2.42(8)   &  0.58(2) \\
   & $(1,1,1)$ &   2.37(21)  &  0.57(5)  \\
   & $(2,0,0)$ &   1.86(48)  &  0.45(12)  \\
\hline
 \end{tabular}
\caption{
The pseudoscalar meson vector form factor from matrix elements of $J_0$ -- fits using Eq.~\protect{\ref{eq:cf}}.
Recall $F(q^2)= 2\kappa Z_V F(q^2)_{latt}$.
The values of $Z_V$ are given in the text.
The momentum $\vec q$ is given in units of $2\pi/L$ or $\pi/8$.
\label{tab:form0}}
\end{table}

\begin{table}
\begin{tabular}{c | c c c}
\hline
 $N_c$ & $q$ &  $V_x$  & $F(q^2)$ \\
 3 & $(1,0,0)$ & 1.12(8)   & 0.68(5)  \\
   & $(1,1,0)$ & 0.96(16)  & 0.58(10)  \\
   & $(1,1,1)$ & 0.88(15)  & 0.53(9)   \\
   & $(2,0,0)$ & 0.99(25)  & 0.30(8)   \\
    \hline
 4 & $(1,0,0)$ & 1.12(6)   & 0.68(4)   \\
   & $(1,1,0)$ & 0.80(5)   & 0.48(3)   \\
   & $(1,1,1)$ & 0.85(13)  & 0.51(8)   \\
   & $(2,0,0)$ & 1.18(41)  & 0.36(12)   \\
\hline
 5  & $(1,0,0)$ & 1.12(6)  & 0.74(4)   \\
    & $(1,1,0)$ & 1.00(5)  & 0.61(3)   \\
    & $(1,1,1)$ & 1.03(7)  & 0.63(4)   \\
    & $(2,0,0)$ & 1.00(29) & 0.31(9)   \\
 \end{tabular}
\caption{
The pseudoscalar meson vector form factor from matrix elements of $J_x$ -- fits using Eq.~\protect{\ref{eq:cmu}}.
Recall $F(q^2)= 2\kappa Z_V V_x/q_x$.
The values of $Z_V$ are given in the text.
The momentum $\vec q$ is given in units of $2\pi/L$ or $\pi/8$.
\label{tab:formx}}
\end{table}

\begin{acknowledgments}
My computer code is based on the publicly available package of the
 MILC collaboration~\cite{MILC}. The version I use was originally developed by Y.~Shamir and
 B.~Svetitsky. Simulations were performed on the University of Colorado Beowulf cluster.
This material is partially
 based upon work supported by the U.S. Department of Energy, Office of Science, Office of
High Energy Physics under Award Number DE-SC-0010005.
\end{acknowledgments}



\end{document}